\documentclass[aps, prb, twocolumn]{revtex4-1}

\usepackage{amsmath,amsfonts,amssymb}
\usepackage{graphicx,color}
\usepackage{hyperref}
\usepackage{epstopdf}
\usepackage{float}
\usepackage{multirow}
\begin{document}

\title{A programmable three-qubit superconducting processor with all-to-all connectivity}

\author{Tanay Roy$^{1}$, Sumeru Hazra$^{1}$, Suman Kundu$^{1}$, Madhavi Chand$^{1}$, Meghan P. Patankar$^{1}$}
\author{R. Vijay$^{1}$}
\email{Corresponding author: r.vijay@tifr.res.in}

\affiliation{$^{1}$Department of Condensed Matter Physics and Materials Science, Tata Institute of Fundamental Research, Homi Bhabha Road, Mumbai 400005, India}

\date{\today}

\begin{abstract}
Superconducting circuits are at the forefront of quantum computing technology because of the unparalleled combination of good coherence, fast gates and flexibility in design parameters. The majority of experiments demonstrating small quantum algorithms in the superconducting architecture have used transmon qubits and transverse qubit-qubit coupling. However, efficient universal digital computing has remained a challenge due to the fact that majority of the state-of-art architectures rely on nearest-neighbor coupling in one or two dimensions. The limited connectivity and the availability of only two-qubit entangling gates result in inefficient implementation of algorithms with reduced fidelity. In this work, we present a programmable three-qubit processor, nicknamed ``trimon", with strong all-to-all coupling and access to native three-qubit gates. We implement three-qubit version of various algorithms, namely Deutsch-Jozsa, Bernstein-Vazirani, Grover's search and the quantum Fourier transform, to demonstrate the performance of our processor. Our results show the potential of the trimon as a building block for larger systems with enhanced qubit-qubit connectivity.

\end{abstract}

\maketitle

Quantum computing offers extraordinary capabilities at solving certain problems by taking classically inaccessible paths. Apart from a robust and scalable architecture, a full-fledged quantum computer will need quantum error correction to prevent irreversible losses arising from decoherence \cite{chuang_book}. Although several proof of principle experiments have been demonstrated in several architectures \cite{qec_NMR,reed_3qubit_qec,qec_catcode,qec_ion}, a fully error-corrected processor has still not been realized.  Therefore, a near-term goal is to build moderate sized processors that can outperform classical computers even without error-correction. Such systems can be realized in an architecture which provides high-fidelity gates to minimize overall errors and maximum inter-qubit connectivity for efficient implementation of algorithms.

A universal quantum computing platform can be constructed using controlled-NOT (or controlled-phase) gate along with single-qubit rotations \cite{Barenco-CNOT}. Nevertheless, interactions beyond two-qubits play an important role in realizing efficient quantum hardware by simplifying complex sequences of two-qubit gates. While specific quantum algorithms have been demonstrated on several quantum computing platforms \cite{DJ-NMR-2qubit-ancilla, NMR-factorize15, DJ-trapped-ion, BV-DJ-optics, Grover-2q-trapped-ion, DJ-BV-2q-dicarlo, DJ-diamond, Shor-trapped-ion, Semiclassical-QFT, Digital-sim-Wallraff, Digitized-adiabaric-Martinis, Martinis-factorize,  Digital-fermion-Martinis}, universally programmable processors have only been built recently\cite{5qubit-ion, Grover-3q-Monroe, 5expt-using-IBM}. However, they have been limited to two-qubit gates native to those systems, requiring a larger sequence of gates and thus limiting their implementation efficiency. One important three-qubit gate that can significantly improve this efficiency is the Toffoli gate \cite{toffoli_book} which flips the state of the target qubit conditioned upon the state of the two control qubits. Together with single-qubit gates, it also forms a universal set for quantum computation \cite{Toffoli-universal}. Although the Toffoli gate has been implemented in various architectures, so far all of them require multiple pulses \cite{qec_NMR, toffoli_wallraff, toffoli_ion_trap, reed_3qubit_qec, dicarlo_GHZ} due to non-availability of a native three-qubit gate.


\begin{figure}[b]
	\centering
	\includegraphics[width=\columnwidth]{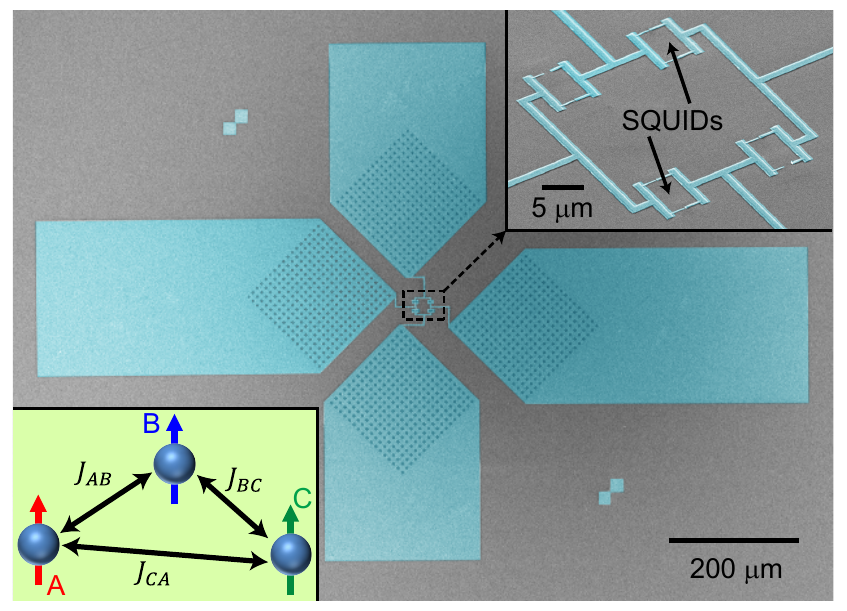}
	\caption{False colored scanning electron micrograph of the trimon device in 3D geomtery. The capacitor pads are placed asymmetrically to achieve optimal level spacing. The waffle pattern close to the junctions is meant for flux trapping. Top-right inset shows a zoomed in view of the junction area where two diagonally opposite junctions are replaced SQUIDs for in-situ tunability of the junction asymmetry. The trimon can be described as a system of three transmon like qubits with strong longitudinal coupling as depicted in the bottom-left inset. }
	\label{fig:trimon}
\end{figure}
Here, we present a three-qubit quantum processor based on multimode superconducting circuits with \textit{all-to-all} connectivity between the qubits. The processor, nicknamed ``trimon", can perform universal quantum computation through execution of a sequence of logic gates completely programmable in software. The native gates in our system are the generalized controlled-controlled-rotations which we achieve with a mean fidelity of about 99\%. We demonstrate a range of three-qubit quantum algorithms using an efficient implementation made possible by access to \textit{error-free} generalized controlled-controlled-phase  gates \cite{multimon}. As examples, Deutsch-Jozsa \cite{DJ_algo}, Bernstein-Vazirani \cite{BV_algo} and Grover's search \cite{Grover_algo} algorithms show success probabilities of 92\%, 63\% and 49\% respectively, significantly larger than the corresponding classical bounds (see Appendix~\ref{app:bounds}) of 50\%, 25\% and 25\%. We also demonstrate the quantum Fourier transform \cite{chuang_book} to estimate phase and amplitude modulation of different three-qubit states. 



Our system consists of a modified version of the previously introduced ``trimon" device \cite{trimon} optimized for three-qubit operations. Scanning electron micrograph of a device nominally identical to the one used in the experiments is shown in Fig.~\ref{fig:trimon}. The capacitor pads are made unequal and are placed asymmetrically to achieve the desired asymmetry for optimal level spacing and coupling strengths \cite{multimon} (see Appendix~\ref{app:dev_par}). The pads are connected to the nodes of a ring with four superconducting arms. Two diagonally opposite arms contain SQUIDs and the remaining two contain single Josephson junctions (top-right inset of Fig.~\ref{fig:trimon}) to control junction asymmetry \textit{in-situ} and tune level spacing. The device gives rise to three coupled anharmonic oscillator modes (labeled as A, B, and C) each of which behaves as a transmon \cite{transmon_theory} qubit. The nonlinearity in the system provides the pairwise inter-mode longitudinal coupling \cite{trimon} (bottom-left inset of Fig.~\ref{fig:trimon}) and the Hamiltonian of the system coupled dispersively to a host readout cavity can be expressed as
\begin{equation}
\label{eq:Hsys}
\begin{split}
\dfrac{1}{\hbar}H_{\rm{disp}} = &\sum_{\mu=A,B,C} \left[ (\omega_\mu - \beta_\mu) \hat{n}_\mu - J_\mu (\hat{n}_\mu)^2 \right] \\
& - \sum_{\mu\neq\nu} 2J_{\mu \nu} \hat{n}_\mu \hat{n}_\nu + \omega_r \left(\hat{n}_r+ \frac{1}{2} \right) \\
& + 2\chi(\hat{n}_A,\hat{n}_B,\hat{n}_C)\hat{n}_r,
\end{split}
\end{equation}
where $\omega_{\mu=A,B,C}$ are the mode frequencies in the absence of any nonlinearity (and hence coupling), $\beta_\mu=J_\mu+ \sum_{\nu\neq\mu}J_{\mu \nu}$ are the shifts due to self-Kerr ($J_\mu$) and cross-Kerr ($J_{\mu \nu}$) type nonlinearities, $\hat{n}_{\mu}$ is the photon number operator for the $\mu$-th mode, $\hat{n}_{r}$ is the cavity's photon number operator, and $\chi$ is the occupation dependent dispersive shift \cite{multimon}.

\begin{figure}[H]
	\centering
	\includegraphics[width=\columnwidth]{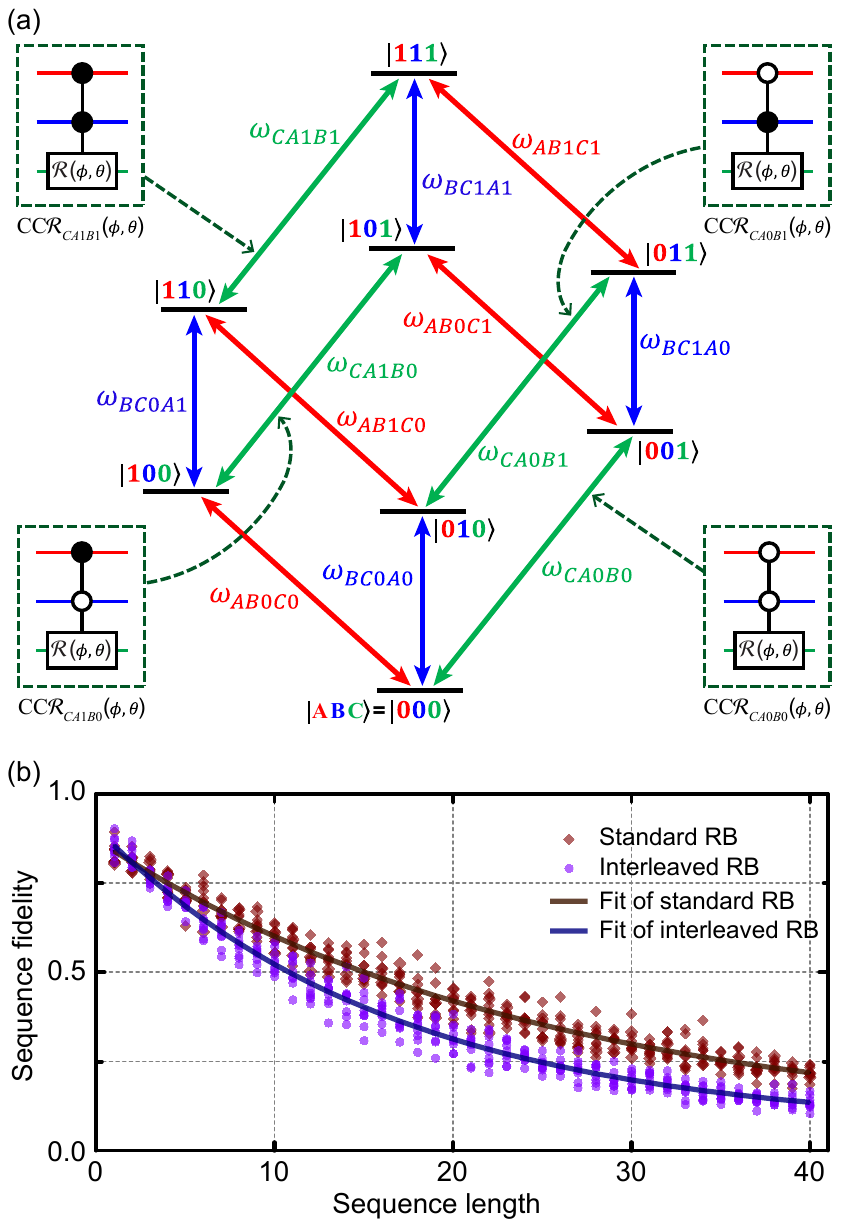}
	\caption{Level diagram and randomized benchmarking. (a) Computational subspace of the trimon consisting of eight energy eigenstates connected through twelve transitions. The trimon can be described as either an eight-level system or a three-qubit system where each qubit has four transition frequencies dependent on the state of its partners. Each transition allows implementation of controlled-controlled-rotation (CC$\mathcal{R}(\phi,\theta)$) through application of a microwave drive. Four CC$\mathcal{R}(\phi,\theta)$ gates for the qubit C are shown inside the dashed boxes. (b) Standard (brown) and interleaved (blue) RB at the transition CA1B1. Each RB sequence is preceded by two $\pi$-pulses at $\omega_{AB0C0}$ and $\omega_{BC0A1}$ to initialize the system in the $|110\rangle$ state. Sequence fidelity is obtained by applying the same $\pi$-pulses in reverse order after the RB sequence and measuring the population in the $|000\rangle$ state. Interleaved RB is performed for the $\pi$-pulse at $\omega_{CA1B1}$ which is equivalent to a Toffoli gate (with appropriate phase advancement). 10 randomized sequences are applied for every sequence length and each data point is a result of averaging 30000 experimental runs. The solid lines correspond to fits from which gate fidelities are extracted. }
	\label{fig:levels_RB}
\end{figure}

The computational subspace of the system is composed of zero and single excitation states of each mode (Fig.~\ref{fig:levels_RB}(a)) so that this subspace can be described as either an eight-dimensional Hilbert space or a composite three-qubit system. The first picture depicts presence of eight energy eigenstates with specific selection rules for the allowed transitions whereas the second picture depicts three qubits whose transition frequencies are dependent on the state of the partner qubits. In what follows, we will stick to the second picture. Each qubit thus has four distinct transition frequencies determined by the inter-qubit coupling strengths, totaling twelve transitions for the whole system. However, one doesn't need individual microwave sources to address all the transitions, since frequencies for a single qubit are within a window narrow enough to be generated comfortably from a single source using sideband modulation techniques.

A microwave drive at each of the twelve transitions implements a controlled-controlled-rotation CC$\mathcal{R}(\phi,\theta)$, where $\phi$ is the azimuthal angle with respect to the $y$-axis and $\theta$ is the polar angle of the Bloch sphere comprising of the two levels connected by the specific transition. The four dashed boxes in Fig.~\ref{fig:levels_RB}(a) show circuit representations for different CC$\mathcal{R}(\phi,\theta)$ gates on qubit C. A $\pi$-pulse about the $x$-axis, CC$\mathcal{R}(-\pi/2,\pi)$, along with appropriate phase shifts of the following pulses \cite{trimon,multimon} then implements a CCNOT or Toffoli gate. Two-qubit controlled-not (CNOT) gates can then be decomposed into two such CCNOT gates and effective single-qubit rotations $\mathcal{R}(\phi,\theta)$ require application of all four CC$\mathcal{R}(\phi,\theta)$ gates. Another interesting feature resulting from the all-to-all longitudinal coupling is the ability to implement \textit{error-free} generalized controlled-controlled-phase (CC$\theta(\theta)$) gate, which is realized by appropriately shifting the phase of all subsequent pulses connecting the particular basis on which the gate is intended to be applied \cite{multimon}. The error-free aspect comes from the fact that the required phase-shifted waveforms are generated in software which can be done with extreme precision (barring any digitization error). Controlled-controlled-Z gate which flips the phase of a particular basis component then becomes a special case: CCZ $=$ CC$\theta(\pi)$. Using the CC$\mathcal{R}(\phi,\theta)$ and CC$\theta(\theta)$ gates, arbitrary three-qubit unitary operation can be implemented on our system.

An experiment on the trimon begins with the determination of the transition frequencies done using standard two-tone spectroscopy with prior one or two $\pi$-pulses for state initialization as one moves up in the ladder(Fig.~\ref{fig:levels_RB}(a)). The TE$_{101}$ mode (with resonance at $\omega_r/2\pi=$ 7.2742 GHz and bandwidth $\kappa/2\pi=3.3$ MHz) of a 3D rectangular copper cavity is used for the joint dispersive measurement \cite{JPA-parity, DJ-BV-2q-dicarlo, joint-readout, joint-readout-2D} of the three qubits. Transitions to the second excited states of each mode are also determined using this technique (while keeping the other two qubits in their ground states) and cross verified by the two-photon transitions from the ground states. Measurement of these fifteen transitions allow determining the anharmonicities ($\alpha_\mu=-2J_\mu$) and inter-qubit coupling ($J_{\mu \nu}$) strengths. Then coherence measurements on the three transitions involving zero to single photon excited states (bottom three transitions in Fig.~\ref{fig:levels_RB}(a): $\omega_{\mu=A,B,C}^{00}$) of each mode are performed to characterize the qubits. Dispersive shifts of the qubits are determined by measuring the shift in cavity frequency when the qubits are excited to the first excited state individually. All the three qubits have well separated frequencies, long coherence times, moderate anharmonicities and strong inter-qubit coupling. These results are summarized in Table.~\ref{table:coherence}.

\setlength{\tabcolsep}{0.95 mm}
\begin{table}[t]
	\caption{Parameters and coherence properties of the trimon device. The transition frequency $\omega^{00}$ of each qubit with the other two qubits in their ground states is listed along with the anharmonicity $(-2J_\mu)$, the relaxation time (T$_1$), the Ramsey decay constant (T$_2^{\rm Ram}$) , the dispersive shift ($\chi$), and the inter-qubit coupling strength ($J_{\mu \nu}$).}
	\begin{tabular}{c c c c c c c} 
		\hline
		\hline
		Qubit & $\omega^{00}/2\pi$ & $-J_\mu/\pi$  & T$_1$ & T$_2^\text{Ram}$ & $\chi/2\pi$  & $J_{\mu \nu}/\pi$ \\
		&(GHz)&(MHz)&($\mu$s)&($\mu$s)&(MHz)&(MHz)\\
		\hline
		A & 4.9278 & -107.9 & 43.7 & 38.9 & -0.064 & $J_{AB}/\pi=192.4$ \\ 
		
		B & 4.5146 & -113.1 & 43.5 & 43.3 & -0.061 & $J_{BC}/\pi=211.4$ \\
		
		C & 5.6864 & -138.0 & 26.3 & 13.6 & -0.069 & $J_{CA}/\pi=242.0$ \\ [0ex]
		\hline
		\hline
	\end{tabular}
	\label{table:coherence}
\end{table}

Next, we characterize our native CC$\mathcal{R}(\phi,\theta)$ gates by performing standard and interleaved randomized benchmarking (RB) \cite{Chow-RB-PRL, Inter-RB}. Generation of these twelve tones are done using three microwave sources where each source is modulated to create four frequencies for a single qubit. Standard RB of each transition involves a sequence of computational gates chosen randomly from the Clifford set addressing the two basis states connected by that transition. Note that these RB experiments should be considered as transition selective RB and not as single-qubit RB (see appendix \ref{app:RB} for details). To determine the fidelity of a particular gate we perform interleaved RB by inserting that gate after every computational gate in a standard RB sequence. Fig.~\ref{fig:levels_RB}(b) show standard and interleaved RB data for the transition $\omega_{CA1B1}$. The average gate fidelity $\mathcal{F}_{\rm avg}$ obtained from standard RB and $\pi$-gate (Toffoli) fidelity $\mathcal{F}_\pi$ obtained from interleaved RB, along with gate lengths, for all twelve transitions are shown in Appendix \ref{app:RB}. The average of $\mathcal{F}_\pi$ across all transitions is found to be $99.1(1)\%$ which is substantially higher than previous demonstrations \cite{toffoli_wallraff, reed_3qubit_qec}. Fidelities of any two-qubit or single-qubit gates can be estimated by taking products of fidelities of the constituent CC$\mathcal{R}(\phi,\theta)$ gates.

\begin{figure}[t]
	\centering
	\includegraphics[width=\columnwidth]{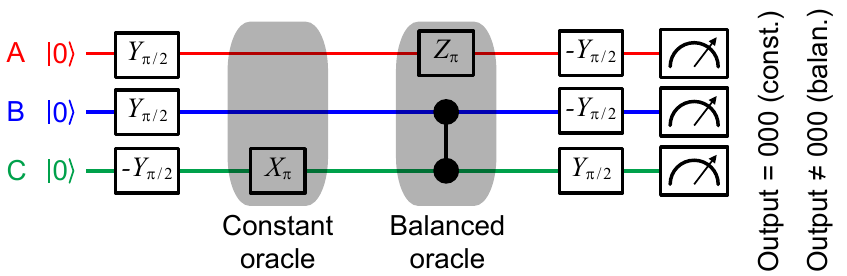}
	\caption{Quantum circuit for the Deutsch-Jozsa (DJ) algorithm. $W_{\theta} (-W_{\theta})$  represents a positive (negative) rotation about the $w$-axis by an amount $\theta$. The oracle is implemented through the application of gates in one of the shaded regions. The constant function is implemented by the presence or absence of the $X_\pi$ gate on qubit C. An example (A$\oplus$BC) for the balanced function is shown. A measurement output of $|000\rangle$ indicates a constant function. }
	\label{fig:DJ}
\end{figure}

\setlength{\tabcolsep}{1.2 mm}
\begin{table*}[t]
	\centering
	\caption{Results of the three-qubit Deutsch-Jozsa algorithm. The success probabilities (SP) for the two implementations of the constant function ($f(x)=0$ or 1) and ten different implementations of the balanced function are shown. The theoretical success probabilities are 100\% for all the cases. The numbers in parentheses represent standard error obtained after 20000 repetitions of each oracle. }
	\label{table:DJ}
	\begin{tabular}{|c||c|c||c|c|c|c|c|c|c|c|c|c|c|}
		\hline
		\multirow{2}{*}{Function} 
		& \multicolumn{2}{c||}{Constant} 
		& \multicolumn{10}{|c|}{Balanced} \\             \cline{2-13}
		& 0 & 1 & A & B & C & A$\oplus$B & B$\oplus$C & C$\oplus$A & A$\oplus$BC & B$\oplus$CA & C$\oplus$AB & AB$\oplus$BC$\oplus$CA  \\  \hline
		SP(\%) & 86.0(3) & 72.6(3) & 93.4(2) & 92.2(2) & 94.9(2) & 97.4(1) & 98.1(1) & 97.5(1) & 96.9(1) & 96.0(1) & 95.0(2) & 93.5(2) \\      \hline
	\end{tabular}
\end{table*}

\begin{figure*}[t]
	\centering
	\includegraphics[width=\textwidth]{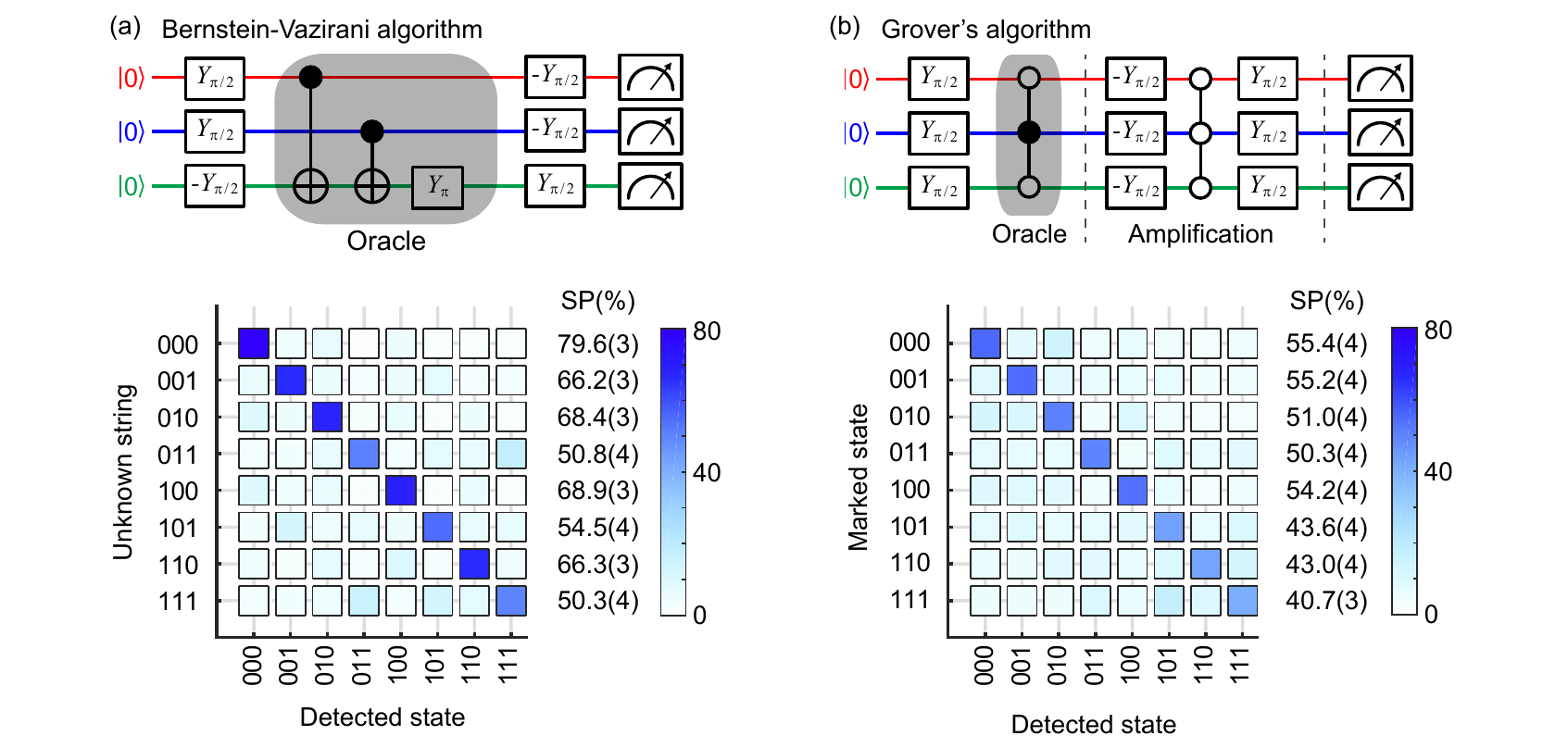}
	\caption{Three-qubit implementations of the Bernstein-Vazirani and Grover's algorithms and their results. (a) The gray area represents the circuit diagram for the oracle implementation. The particular oracle is showing the realization of the unknown string `111', while the string `000' would be realized by application of no gates. The bottom panel shows experimental outcomes indicating success probabilities (SP). The theoretically expected SPs are 100\% for all cases. (b) The oracle for Grover's algorithm is realized by the application of CCZ gates (connected black/white circles; black representing 1 and white representing 0) conditioned on different qubit states. The probability amplification is performed by a CCZ$_{000}$ gate (which flips the phase of $|000\rangle$) sandwiched between single-qubit $-\pi/2$ and $+\pi/2$ rotations about the $y$-axis on all the qubits. The bottom panels show SPs for each marked state. Ideal values for a single iteration are 78.125\%. Each of the oracle implementations are repeated for 20000 times for both algorithms. The numbers in parentheses represent the standard error of the outcomes. }
	\label{fig:BV_Grover}
\end{figure*}

In order to demonstrate the universal programmability of our processor and characterize its performance, we run various three-qubit quantum algorithms. We start with the Deutsch-Jozsa (DJ) algorithm \cite{DJ_algo}, a simple quantum algorithm to show quantum advantage over classical methods. The DJ algorithm determines whether a function $f$ (also called the ``oracle") is balanced or constant. The function takes an $n$-bit register $\boldsymbol{x}$ as input and is said to be balanced if the 1-bit output is 0 for exactly half of the possible inputs and 1 for the rest, whereas $f$ is called a constant function if the output is always 0 or 1 irrespective of the input. In the classical case one needs 2 (best scenario) to $2^{n-1}+1$ (worst scenario) evaluations to determine the nature of the function, whereas the quantum algorithm requires only one evaluation. 

Fig.~\ref{fig:DJ} shows the quantum circuit for our (\textit{ancilla-free}) implementation of the DJ algorithm. The algorithm starts with preparation of a specific equal superposition state $(|0\rangle+|1\rangle)(|0\rangle+|1\rangle)(|0\rangle-|1\rangle)/2\sqrt{2}$ using $\pm Y_{\pi/2}$ gates (positive or negative rotation about the $y$-axis by an amount $\pi/2$ which are composed of four CC$\mathcal{R}(0,\pi/2)$ pulses). We implement the constant function by either applying or not applying an $X_{\pi}$ gate ($\pi$ rotation about the $x$-axis) on qubit C which corresponds to $f(x)=1$ and 0 respectively (see Fig.~\ref{fig:DJ}). Ten different kinds of balanced functions are implemented by applying a combination of $Z_\pi$ and controlled-Z gates. After application of one of the oracles, the states are rotated back before the final measurement. A final outcome of $|000\rangle$ indicates a constant function whereas anything else implies a balanced function. We obtained mean success probabilities (SP) of 95.5(1)\% and 79.3(3)\% for the balanced and constant functions respectively with theoretical predictions being 100\% for both cases. None of the algorithmic outcomes are corrected for state preparation and measurement (SPAM) errors. Individual success probabilities for all the functions are tabulated in Table~\ref{table:DJ} and are comparable with previous demonstrations \cite{NMR-3qubit-DJ, 5qubit-ion}.

The Bernstein-Vazirani (BV) \cite{BV_algo} algorithm is similar to the DJ algorithm, where the goal is to determine an unknown string of binaries $\boldsymbol{c}=\{c_1,c_2,\cdots, c_n\}$. The oracle is defined as $\boldsymbol{c\cdot x}$ mod 2, where $\boldsymbol{x}$ is the input register to which the user has full control. The BV algorithm just takes one execution of the oracle to find the unknown string $\boldsymbol{c}$. The quantum circuit and results are shown in Fig.~\ref{fig:BV_Grover}(a). The eight instances of the oracle are implemented by the application of conventional CNOT$_{A\rightarrow C}$, CNOT$_{B \rightarrow C}$ (each composed of two CCNOT) and $Y_\pi$ (composed of four CC$\mathcal{R}(0,\pi)$) gates on qubit C conditioned on the classical bits being 1 in the unknown string. The results are shown in Fig.~\ref{fig:BV_Grover}(a). On average, correct results are obtained with a probability of 63.1(3)\% which is not as good as previous demonstrations in optical and trapped-ion systems \cite{BV-DJ-optics, 5qubit-ion}. However, ours is the first demonstration of a three-qubit BV in a solid-state system with further possibility of improvement as discussed later.

\begin{figure*}[t]
	\centering
	\includegraphics[width=\textwidth]{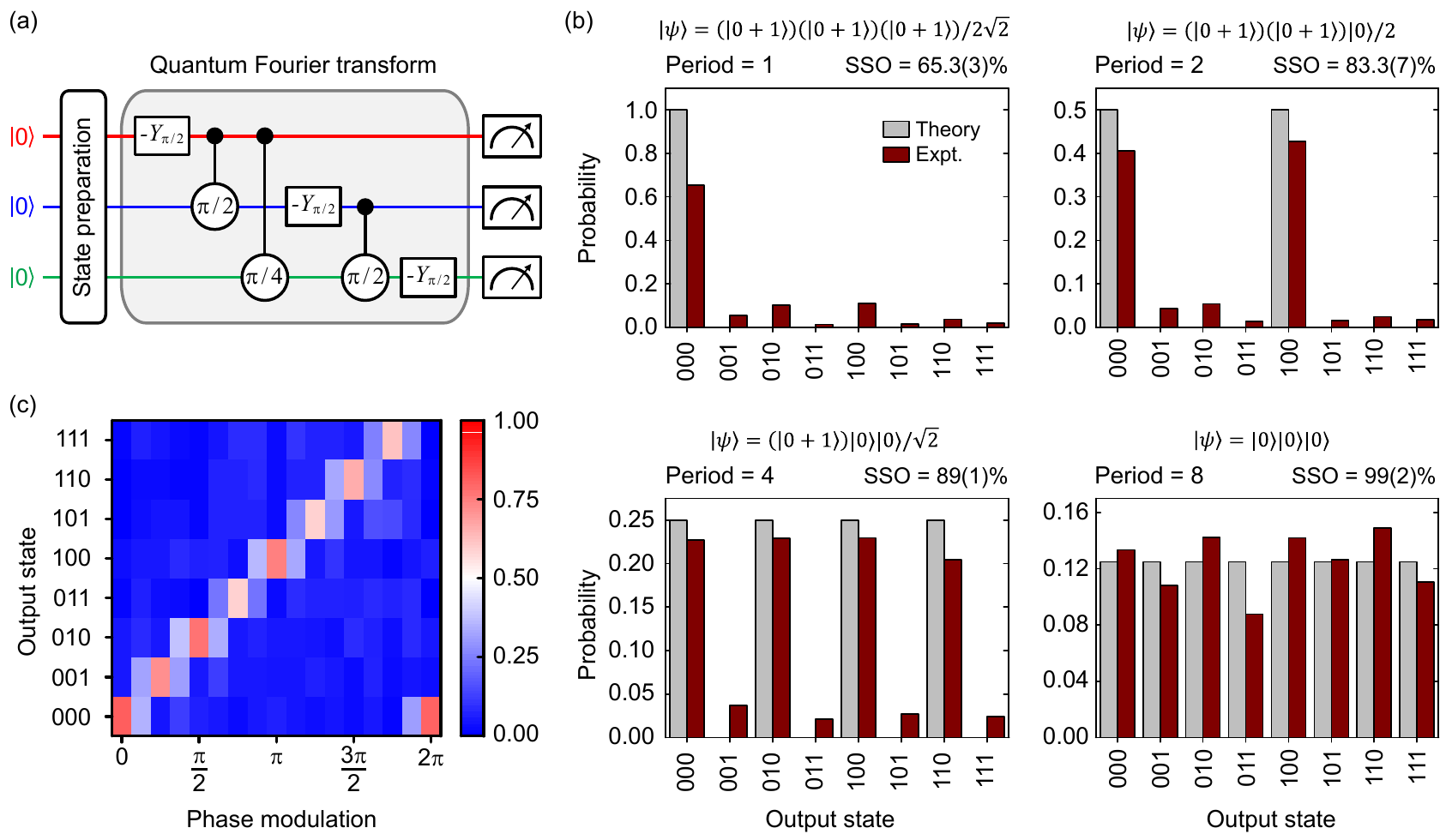}
	\caption{Three-qubit quantum Fourier transform (QFT) protocol and results. (a) The quantum circuit for performing QFT. The state preparation step consists of single-qubit rotations to achieve desired periodicity in the coefficients $C_j$ in $|\psi\rangle=\sum_{j=0}^7C_j|j\rangle.$ The gray area represents the implementation of QFT circuit using twelve native pulses (for -$Y_{\pi/2}$ gates) and three virtual CC$\theta$ gates. The registers are finally measured to obtain the state populations. (b) Results of the period-finding protocol for four different initial states with periodicities 1,2,4 and 8. The gray and brown bars represent theoretical and experimentally obtained values. The squared statistical overlap (SSO) denote the fidelity of the process with statistical uncertainties obtained from 20000 repetitions. (c) Results of the phase estimation protocol. Modulating phase is swept over a range of 0 to $2\pi$ in steps of $2\pi/16$. Mean fidelity obtained is 81.3(9)\%.}
	\label{fig:QFT}
\end{figure*}

The DJ and BV algorithms are deterministic in nature, however, there are certain quantum algorithms which find the correct answer probabilistically and yet provide quantum advantage. Next, we demonstrate one such algorithm, namely the Grover's algorithm \cite{Grover_algo}, which can provide quadratic speedup in performing search of an unsorted database. This algorithm can find, in $O(\sqrt{N})$ steps, the unique input that produces a specific output when fed to a database of size $N$. We perform a single iteration of Grover's search on our three-qubit system with database size $N=2^3=8$. The theoretical probability of getting the right answer after a single iteration of Grover's algorithm for $N=8$ is 78.125\% \cite{Grover_bound}, compared to 25\% for the classical case\cite{Grover-3q-Monroe}. Our implementation of the quantum circuit is shown in Fig.~\ref{fig:BV_Grover}(b) which does not require any ancillary qubit. The algorithm starts with preparation of the equal superposition state $(|0\rangle+|1\rangle)(|0\rangle+|1\rangle)(|0\rangle+|1\rangle)/2\sqrt{2}$, realized by $Y_{\pi/2}$ gates on all qubits, followed by the mapping of the marked state through an oracle which we implement using appropriate CCZ gates. The oracle flips the probability amplitude of the marked state which is then amplified by the next stage of operations (gates within the dashed lines in Fig.~\ref{fig:BV_Grover}(b)). Finally, the qubits are measured to find the answer. Note that our circuit is significantly simpler than the conventional circuit \cite{chuang_book, Grover-3q-Monroe}, because of access to generalized CCZ gates. The outcomes for the eight possible oracles are shown in Fig.~\ref{fig:BV_Grover}(b) with mean SP of 49.2(4)\%. These results are on par with a recent demonstration in the trapped-ion system \cite{Grover-trapped-ion}, although much better results were expected due to access to native three-qubit gates. The first limiting factor is the measurement error which is discussed in detail later. The other fact is the requirement of preparation of equal superposition state, which is currently implemented inefficiently in our system.

Finally, we implement Quantum Fourier transform (QFT) which is a key ingredient in many quantum algorithms that exploit parallel functional evaluation through preparation of a superposition state to achieve exponential speedup over classical algorithms \cite{chuang_book}. Two such algorithms are \textit{order-finding} and \textit{phase estimation}. Order-finding protocol plays a crucial role in Shor's factorizing algorithm \cite{Shor-algo} and the phase estimation protocol determines the phase $\phi$ of an eigenvalue problem $\hat{A}|\phi\rangle=e^{i\phi}|\phi\rangle$ with $n$-bit precision using an $n$-qubit QFT \cite{chuang_book}.

The quantum circuit for the three-qubit QFT is shown in Fig.~\ref{fig:QFT}(a) consisting of twelve native gates (each single-qubit gate composed of four CC$\mathcal{R}$ gates) and three virtual CC$\theta$ gates. Note that the Hadamard gates are replaced with $-Y_{\pi/2}$ gates which lead to identical projections on subsequent measurement and the swap between qubits A and C are done in software after recording the outcomes. To demonstrate the period-finding protocol, we initialize the system in a superposition state of the form $|\psi\rangle=\sum_{j=0}^7 C_j|j\rangle$, where coefficients $C_j$ are chosen to exhibit amplitude periodicities 1,2,4 and 8 (see Fig.~\ref{fig:QFT}(b)). After performing the QFT, amplitude modulation appears in the distribution of state populations. To compare the results with the theoretical predictions, we define squared statistical overlap (SSO) \cite{Semiclassical-QFT} as SSO $=\left(\sum_{j=0}^{N-1}\sqrt{t_j e_j} \right)^2$, where $t_j$ and $e_j$ are the theoretically expected and experimentally obtained probabilities respectively for state $|j\rangle$. The SSOs for each input state are shown in Fig.~\ref{fig:QFT}(b), with an average of 84(1)\%.

The phase estimation protocol is identical to that of the period-finding where the input state is prepared in the form $|\psi\rangle = \frac{1}{2\sqrt2}\bigotimes_{k=0}^2 \left( |0\rangle + e^{-i2^k \phi}|1\rangle \right)$. Preparation of these states requires rotation of each qubit about $Z$ axis and $|\psi\rangle$ exhibits $\phi$ dependent phase modulation in the coefficients $C_k$. Performing QFT on this state allows determination of $\phi$ with a 3-bit precision which is mapped on to the population of the states. We swept $\phi$ from 0 to $2\pi$ in steps of $2\pi/16$ and the results are shown in Fig.~\ref{fig:QFT}(c). Values of $\phi$ that are integer multiples of $2\pi/8$ result in the state $|8\phi/2\pi\rangle$ after the QFT, whereas for non-integer multiples the value of $\phi$ is mapped on to nearest 3-bit states. The average success probability for the cases of integer multiples is 70.1(3)\% and the mean SSO for the non-integer multiples is 92(1)\%. These results are again similar to the state-of-art implementation in the trapped-ion system \cite{5qubit-ion}.


Our results are mainly limited by measurement error (mean error $10.5\%$) due to relatively large overlaps between histograms (see Fig.~\ref{fig:histogram} in Appendix). The reason behind our inability to improve the measurement error is small dispersive shifts of the qubits for the specific sample (Table~\ref{table:coherence}). Dispersive shifts can be easily increased by several times through modifying the geometry of the capacitor pads while retaining optimal level spacing. Another way of improving the results is to reduce evaluation time of the algorithms. The processor is programmed to apply the (native CC$\mathcal{R}$) gates sequentially, but all the pulses for a single qubit can be applied simultaneously by generating multi-frequency pulses and that will reduce the execution time significantly. One advantage we achieve by making the trimon asymmetric is that all the three modes are now strongly coupled, leading to negligible ac Stark shifts. Given the separation between different transitions, we can therefore easily increase Rabi rates by a factor of 2 to 3 or even more by appropriate pulse shaping and implementing a full derivative removal by adiabatic gate (DRAG) pulse \cite{drag-pulse}. Another problem we have noticed (among several samples), that reduce the rate of data acquisition, is the occasional jumps in the system parameters and instability in one of the qubits. Understanding this instability and implementing possible modifications described above are the subject of immediate future work.

In summary, we have presented a universally programmable solid-state three-qubit processor with unparalleled performance in superconducting architecture. Our results clearly illustrate the flexibility of the system and novel ways of efficiently implementing various algorithms. The efficiency originates from the all-to-all connectivity and access to native three-qubit gates. These two features enabled us to demonstrate the ancilla-free versions of the three quantum algorithms \cite{NMR-3qubit-DJ} and thus cutting down the required number of physical qubits \cite{5qubit-ion}. We envisage the trimon as a building block for larger systems where multiple trimons are coupled to a common bus resonator \cite{multimon} and inter-trimon gates are performed by adapting the well established techniques of cross-resonance \cite{CR-gate}. This will help in realizing a small-scale processor with enhanced connectivity resulting in more efficient implementation of algorithms.

\begin{acknowledgements}
This work is supported by the Department of Atomic Energy of the Government of India. R.V. acknowledges support from the Department of Science and Technology, India via the Ramanujan Fellowship and Nanomisson. We thank Anirban Bhattacharjee for help with numerical simulations and Pranab Sen for useful discussions. We acknowledge the TIFR nanofabrication facility.

\textbf{Author contributions.}
T.R. and S.H. designed the device. S.H., M.C and M.P.P. fabricated the device. T.R. designed the experimental protocols and performed simulations. T.R. and S.K. performed the experiments and T.R. analyzed the data. S.H. developed the MLE code for density matrix reconstruction. T.R., S.H. and R.V. wrote the manuscript with inputs from co-authors. R.V. supervised the whole project.
\end{acknowledgements}

\appendix

\section{Classical bounds for different quantum algorithms}
\label{app:bounds}

We would like to compute the maximum classical success rates for problems addressed by Deutsch-Jozsa (DJ) \cite{DJ_algo}, Bernstein-Vazirani (BV) \cite{BV_algo}, and Grover's \cite{Grover_algo} algorithm for one iteration of the functional evaluation. In the first case, given a specific output, the chance of the function being balanced or constant is exactly half. Therefore, the best strategy is to guess the function randomly, resulting in a success rate of 50\%.

A single functional evaluation using a classical circuit for the determination of an unknown string $\boldsymbol{c}$ can, at maximum, reveal one bit of information. Hence, if $\boldsymbol{c}$ is a string of three bits, the other two bits have to be chosen randomly. Therefore, the probability of success in this case is $\frac{1}{2}\cdot\frac{1}{2}=25$\%.

The optimal classical strategy for searching an unsorted database (of size $2^3=8$) provides the correct answer with a probability of $\frac{1}{8}+\frac{7}{8}\cdot\frac{1}{7}=25$\%. This result corresponds to a chance of $\frac{1}{8}$ getting the right outcome in the first iteration, followed by a random guess in case the first query failed.

\section{Device Parameters}
\label{app:dev_par}

It is important to choose the device parameters in such a way that all transitions in the level diagram (see Fig.~\ref{fig:levels_RB}(a)) are well separated from each other for fast gate operations and prevention of leakage out of logical subspace. Another important criterion is to ensure that the absolute energy of the $|111\rangle$ levels is significantly smaller than the available potential height of 4$E_{J_{\rm min}}$ set by the smallest of the four junctions \cite{multimon}. In order to find the required device parameters, we first start with a set of guess parameters and a target level diagram where each transition is separated by at least 30 MHz from any other transition. Then we run a minimization routine (downhill simplex) using the method described in Ref.~\onlinecite{multimon} to find the Josephson junction energies and capacitances constrained to feasible values. A finite element simulation (using COMSOL\textsuperscript{\textregistered}) is used to find the geometric structure that gives rise to the required capacitance values. Josephson energies for different junctions are converted into junction overlap dimensions from independent fabrication calibrations. Finally, the device is prepared using standard lithographic and electron beam evaporation techniques. For the device presented here, required asymmetries are put in the capacitor structure while keeping the junctions identical to minimize fabrication uncertainties which are prominent for smaller features. Further, two of the diagonally opposite junctions are replaced with SQUIDs (see Fig.~\ref{fig:trimon}) for \textit{in-situ} control of the junctions asymmetry to play with the level spacings.

After experimental determination of the frequencies, inter-qubit coupling strengths, and anharmonicities, one can use the same minimization tool to predict the device parameters that would have led to the properties observed. For the device presented here, the best fit parameters are: $E_{J_{12}}=7.49$ GHz, $E_{J_{23}}=6.88$ GHz, $E_{J_{34}}=7.40$ GHz, $E_{J_{41}}=6.43$ GHz, $C_{12}=37.95$ fF, $C_{23}=34.30$ fF, $C_{34}=32.91$ fF, $C_{41}=27.53$ fF, $C_{13}=10.84$ fF, $C_{24}=19.19$ fF, where the nodes are numbered anti-clockwise starting from the top pad. Ground capacitances are assumed to be 0.01 and 0.02 fF for the smaller and larger pads respectively. To achieve the optimal level spacing we operate at a flux-bias which corresponds to integer multiple of the flux-quantum for the ring but fractional flux-quantum for the small squid loops. This operating point is evident from the observation that bottom-left ($E_{J_{23}}$) and top-right ($E_{J_{41}}$) junctions have smaller Josephson energies than the rest. The remaining asymmetries in the junction energies can be attributed to fabrication uncertainties ($\sim 3\%$). The capacitance values are in good agreement ($\sim 3\%$ mismatch) with the estimates from the COMSOL\textsuperscript{\textregistered} simulation.

\section{Signal Generation and Measurement Setup}
\label{app:setup}

\begin{figure*}[t]
	\centering
	\includegraphics[width=\textwidth]{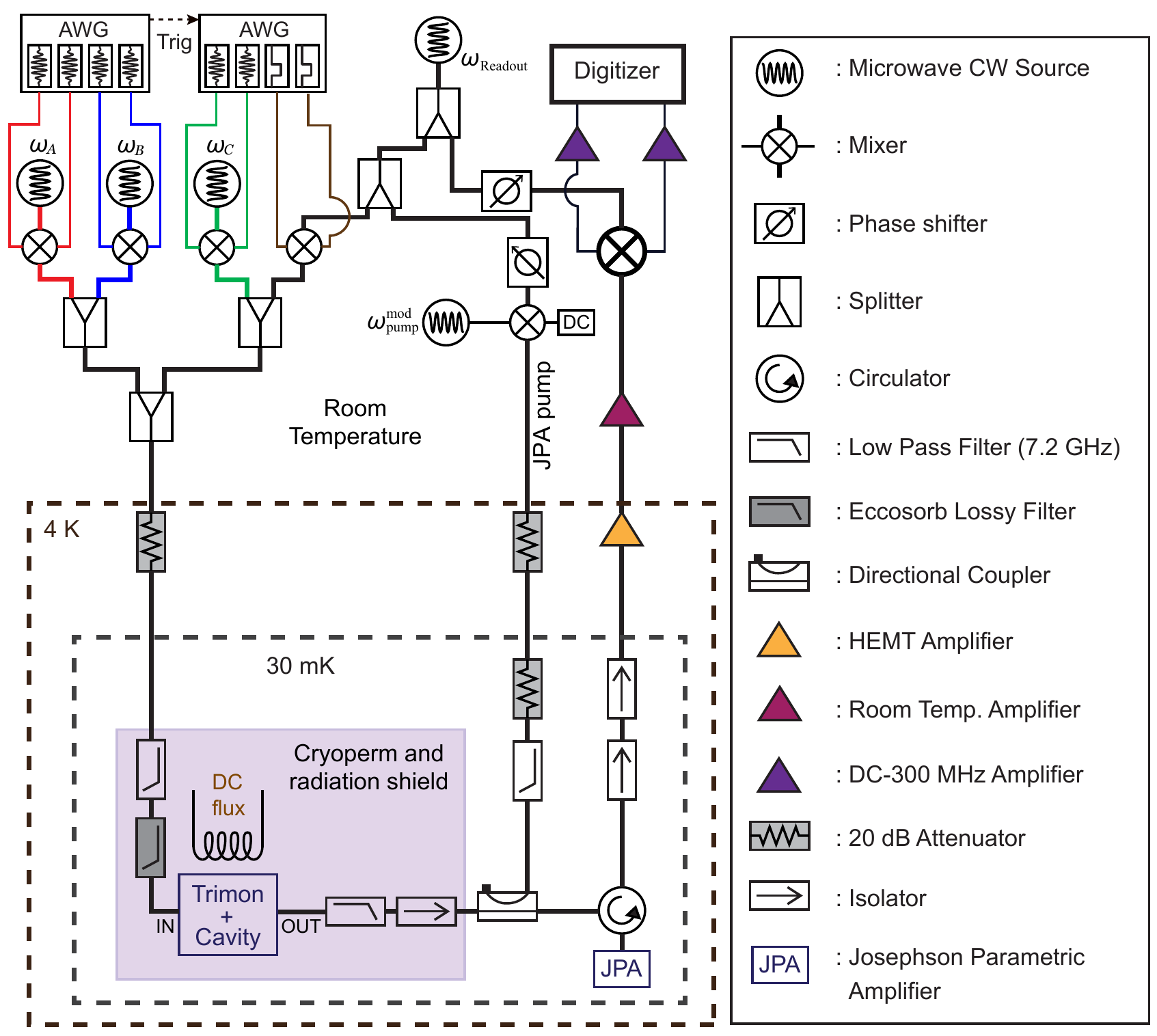}
	\caption{Room temperature and cryogenic circuitry for the three-qubit measurement setup.}
	\label{fig:setup}
\end{figure*}
The detailed measurement setup is shown in Fig.~\ref{fig:setup}. We use four microwave sources - three for the qubits and one for the readout. Pulses at the four different frequencies for a single qubit are generated by modulating a microwave tone kept close to the mean of two extreme transition frequencies. Two channels of an arbitrary waveform generator (AWG) running at $2\times10^9$ samples/s are used for the sideband modulation of each qubit. We use two 4-channel AWGs, one of them being triggered by the master AWG. Modulating waveforms sent to two AWGs are generated with appropriate time shifts to account for the trigger delay and synchronize the outputs. Input signals are passed through a 20-dB attenuator (at 4 K stage) followed by reflective low-pass and lossy Eccosorb\textsuperscript{\textregistered} filters \cite{slichter-filter} attached to the base plate. The trimon device is placed inside a copper 3D rectangular cavity with resonant frequency $\omega_r/2\pi=$7.2742 GHz (TE$_{101}$ mode). Qubit state preparation and joint dispersive readout are performed by sending microwave tones through a common input coaxial line. The transmitted signal at $\omega_r$ is first amplified using a near-quantum-limited Josephson parametric amplifiers \cite{Hatridge-JPA} followed by a commercial high-electron-mobility-transistor (HEMT) amplifier at 4 K and room temperature amplifiers. The homodyne detection of the outgoing signal is performed by demodulation, further amplification and digitization at a rate of $100\times10^6$ samples/s. The digitized signal is stored and processed inside a computer.

\section{Randomized Benchmarking}
\label{app:RB}

We estimate the gate fidelity corresponding to a particular transition by carrying out Clifford based standard randomized benchmarking (RB) in the two-level subspace corresponding to that transition. In this protocol we choose a group of rotations that maps the two level system between two polar and four equally spaced equatorial points on the Bloch sphere. We arrange them in a random fashion to remove any bias from the gate error. Finally we apply a recovery pulse which is inverse of the effective rotation due to all previous gates of that sequence. We do a projective measurement on the final state. The initial state population at the end of the sequence is defined as the sequence fidelity. As we vary the length ($N_s$) of the sequences, error accumulation gives rise to an exponential decay of the sequence fidelity $\mathcal{F}_{s}$ which follows as $\mathcal{F}_{s}=Ap_s^{N_s}+B$. The SPAM error affects only $A$ and $B$ and leaves the decay rate $p_s$ unaffected. The average gate fidelity $\mathcal{F}_{\rm avg}$ is then obtained from $\mathcal{F}_{\rm avg}=(1+p_s)/2$.

\setlength{\tabcolsep}{1.1 mm}
\begin{table}[t]
	\label{table:RB}
	\caption{Average gate fidelities $\mathcal{F}_{\rm{avg}}$ and generalized Toffoli gate fidelities $\mathcal{F}_\pi$ at the twelve transitions obtained from standard and interleaved randomized benchmarking experiments respectively. The numbers in parentheses represent fitting error. Transition frequencies and $\pi$-pulse lengths are also tabulated.}
	\centering
	\begin{tabular}{c c c c c} 
		\\ [-2 ex]
		\hline
		\hline
		Transition & $\omega/2\pi$ (GHz) & $\mathcal{F}_{\rm{avg}}$  & $\mathcal{F}_\pi$ & $\pi$-pulse (ns) \\
		\hline
		AB0C0 & 4.9278 & 0.994(2) & 0.999(2) & 125 \\
		
		AB0C1 & 4.6858 & 0.988(2) & 0.989(1) & 226\\
		
		AB1C0 & 4.7355 & 0.993(1) & 0.993(1) & 230 \\ 
		
		AB1C1 & 4.4837 & 0.978(1) & 0.985(1) & 316 \\
		
		BC0A0 & 4.5146 & 0.992(2) & 0.996(1) & 617 \\
		
		BC0A1 & 4.3222 & 0.989(1) & 0.995(1) & 257 \\
		
		BC1A0 & 4.3032 & 0.981(1) & 0.983(1) & 293 \\
		
		BC1A1 & 4.1011 & 0.980(1) & 0.984(1) & 333 \\
		
		CA0B0 & 5.6864 & 0.992(1) & 0.992(1) & 319 \\
		
		CA0B1 & 5.4750 & 0.989(1) & 0.994(1) & 311 \\
		
		CA1B0 & 5.4444 & 0.986(1) & 0.992(1) & 295 \\ 
		
		CA1B1 & 5.2232 & 0.982(1) & 0.986(1) & 333 \\
		[0ex]
		\hline
		\hline
	\end{tabular}
\end{table}

To quantify the error in a particular gate $\mathcal{G}$, we first obtain a reference curve by performing standard RB and extract the decay rate $p_{\rm{ref}}$. Now we interleave the target gate $\mathcal{G}$ between the randomly generated Cliffords and fit the decay curve to another exponential of the same form resulting in a decay rate $p_{\mathcal{G}}$. The error of the target gate can be estimated by separating out the error due to the reference RB using the formula, $r_{\mathcal{G}}=(1-p_{\mathcal{G}}/p_{\rm{ref}})/2$. Fidelity $\mathcal{F}_\mathcal{G}$ of the gate $\mathcal{G}$ is then obtained from $\mathcal{F}_\mathcal{G}=1-r_{\mathcal{G}}$.

In order to execute RB addressing a particular transition in the full three-qubit Hilbert space, we first initialize the system to the lower energy state of that particular two level subspace. This might require zero, one or two additional CCNOT pulses depending on the particular transition. Every RB sequence is appended with pulses that will bring the system back to $|000\rangle$ before the final projective measurement. The sequence fidelity is then measured by the population of the $|000\rangle$ state. For every RB experiment we have varied $N_s$ from 1 to 40 and for every given length 10 random Clifford sequences are chosen. The  fidelity of each such sequence is estimated by performing 30000 repetitions. The average Clifford gate fidelities and the generalized Toffoli (CC$\mathcal{R}(-\pi/2,\pi)$) gate fidelities obtained from standard and interleaved RB for all the transitions are tabulated in Table~III.

\section{Qubit Readout and State Tomography}
\label{app:rdt}

\begin{figure}[t]
	\centering
	\includegraphics[width=\columnwidth]{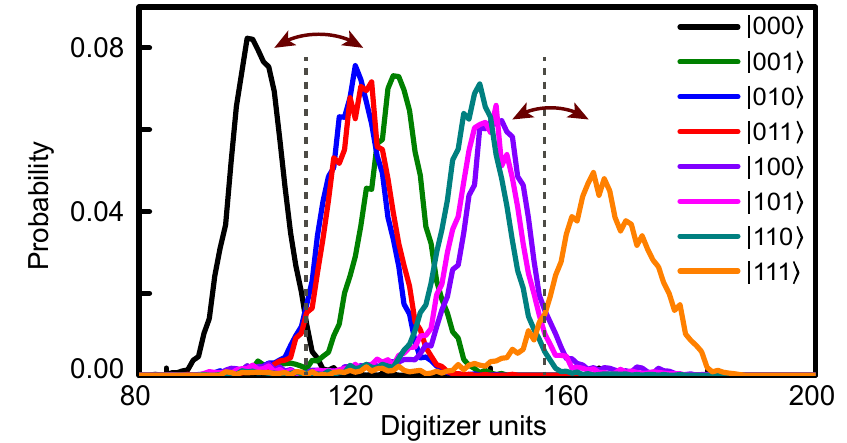}
	\caption{Typical histograms obtained from the $\sigma_z$-measurement of the basis states for a trimon. Two demarcation (dashed gray) lines can be drawn to distinguish states $|000\rangle$ and $|111\rangle$ from the rest. Then appropriate CCNOT gates are applied to find projections along other basis directions.}
	\label{fig:histogram}
\end{figure}

Since all the three qubits share the same host cavity, they are subject to joint dispersive readout \cite{JPA-parity, DJ-BV-2q-dicarlo, joint-readout, joint-readout-2D}. Typical histograms of all the basis states resulting from a $\sigma_z$-measurement are shown in Fig.~\ref{fig:histogram}. Although large overlaps between some basis states appear because of similar dispersive shifts (see Table~\ref{table:coherence}), states $|000\rangle$ and $|111\rangle$ have a very small overlap with the rest and can be measured with high confidence. We draw two demarcation lines (dashed gray lines in Fig.~\ref{fig:histogram}) to separate states $|000\rangle$ and $|111\rangle$ from the rest, and discard any outcomes which fall between the two lines. Thus, in the first measurement, one finds projections along $|000\rangle$ and $|111\rangle$. Then to find projections along $|001\rangle$ and $|110\rangle$ two CCNOT gates are applied at frequencies $\omega_{CA0B0}$ and $\omega_{CA1B1}$ to exchange population between pairs $|000\rangle \leftrightarrow |001\rangle$ and $|110\rangle \leftrightarrow |111\rangle$. In the next iteration, two more CCNOT gates are applied to perform measurements along  $|010\rangle$ and $|101\rangle$ and so on. One thus needs four rounds of measurements to find projections of all three-qubits along the $\sigma_z$ direction. Measurement fidelities for the states $|000\rangle$ and $|111\rangle$ are $\mathcal{F}_{000}=$ 95.1(2)\% and $\mathcal{F}_{111}=$ 85.2(3)\% respectively obtained from the extreme two histograms in Fig.~\ref{fig:histogram}. Measurement fidelities for the other basis states are obtained by multiplying these fidelities with the fidelity of the relevant CCNOT gate used for the swap operation.

\section{Tomography and fidelity of various states}
\label{app:states}

\begin{table*}[t]
	\label{table:fidelity2}
	\caption{State fidelities obtained experimentally ($\mathcal{F}_{\rm expt}$) and from numerical simulation ($\mathcal{F}_{\rm sim}$) of various two-, three-qubit entangled and equal superposition states along with the number of pulses required for preparation. The numbers in parentheses represent fidelity error obtained from bootstrapping.}
	\begin{tabular}{c c c c c}
		\\
		\hline
		\hline
		Name & State & Pulses & $\mathcal{F}$(\%) & $\mathcal{F}_{\rm sim}$(\%) \\
		\hline \\ [-2 ex]
		Bell & $\dfrac{|000\rangle +  |110\rangle}{\sqrt{2}}$ & 2 & 95.3(1) & 96.7\\ [2 ex]
		GHZ & $\dfrac{|000\rangle + i|111\rangle}{\sqrt{2}}$ & 3 & 93.4(1) & 95.1\\ [2 ex]
		Werner & $\dfrac{|100\rangle + |010\rangle + |001\rangle}{\sqrt{3}}$ & 3 & 95.4(1) & 96.5 \\ [2 ex]
		Equal superposition & $\dfrac{(|0\rangle + |1\rangle)^{\otimes3} }{2\sqrt{2}}$ & 7 & 93.3(1) & 95.9 \\ [2 ex]
		\hline
		\hline
	\end{tabular}
\end{table*}

Apart from the quantum algorithms presented in the main text, we have prepared various entangled and product states and performed three-qubit tomography for verification. Performing tomography on an arbitrary state requires measurement along $x,y,z$-axes of all qubits, i.e., along $3\times3\times3=27$ orthogonal directions. This is done by applying all combinations of pre-rotations on individual qubits along $x$ or $y$-axes before performing the $\sigma_z$-measurement to find all the necessary projections. Then these projections are used to reconstruct the most probable density matrix $\rho_{\rm{expt}}$ using the maximum likelihood estimation technique \cite{MLE,MLE1,MLE2}. The fidelity of the reconstructed state is calculated using the definition $\mathcal{F}_{\rm expt}=$Tr$\left[ \sqrt{\sqrt{\rho_{\rm{th}}} \rho_{\rm{expt}} \sqrt{\rho_{\rm{th}}}} \right]$, where $\rho_{\rm{th}}$ is the ideal density matrix corresponding to the state. The fidelities of the states prepared, along with a number of native pulses required for preparation are tabulated in Table~IV. These fidelity numbers are corrected for measurement errors due to histogram overlaps. In order to characterize the performance of our processor, we also performed numerical simulations using the QuTiP open-source software package\cite{qutip1,qutip2} with the decoherence parameters shown in Table~\ref{table:coherence}. The fidelities $\mathcal{F}_{\rm sim}$ are very close to what we obtained experimentally. The small discrepancy can be attributed to measurement errors which are not included in the simulations.

%

\end{document}